\documentclass{article}
\usepackage{amsmath}
\usepackage{amsthm}
\usepackage{url}
\usepackage{amsfonts}
\usepackage{carl}
\usepackage[vlined,linesnumbered]{carlalgorithm}
\usepackage[latin1]{inputenc}

\title{A note on the security of the \textit{h}HB protocol}
\author{Carl L\"ondahl\\~\\\texttt{carl@grocid.net}}
\begin{document}
\maketitle
\begin{abstract} We propose a polynomial-time attack on the \textit{h}HB protocol, showing that the protocol does not attain the security it claims. Our attack is based on the attack introduced in \cite{Gilbert05}.\end{abstract}
\section{Introduction}
In the modern era of cryptography, researchers have struggled with finding low-cost cryptographic primitives suitable for simplistic hardware environments such as RFID tags and low-power/cost devices. A popular source of inspiration is the \emph{Learning Parity with Noise} (\textprob{LPN}) problem, which has roots in machine learning theory but now has gained a lot of popularity among cryptographers. The \textprob{LPN} problem is strongly related to the problem of \textit{decoding random linear codes}, which probably is the most important problem in coding theory. Being supposedly hard, \textprob{LPN} plays an important role in post-quantum cryptography in contrast to classic number theoretic problems.  \textprob{LPN} consists of very basic arithmetic operations and is therefore a perfect fit for light-weight cryptography.

The first 'real' cryptographic construction based on \textprob{LPN} was the Hopper-Blum (HB) protocol \cite{HB} -- a minimalistic protocol being secure in a \emph{passive} attack model. Juels and Weis \cite{JW}, and Katz and Shin \cite{KatzShin} proposed a modified protocol, HB$^+$, which aimed to be secure also in the \emph{active} attack model by extending HB with one extra round.
However, Gilbert \textit{et al.} \cite{Gilbert05} later showed that the HB$^+$ protocol is vulnerable to active attacks, {i.e.}, man-in-the-middle attacks. Later, Gilbert \textit{et al.} \cite{Gilbert08} proposed a variant of the Hopper-Blum protocol called HB$^\#$.

Some of the more recent contributions to \textprob{LPN}-based constructions are a
two-round identification protocol called Lapin, proposed by Heyse \textit{et al.} \cite{Lapin}, and an LPN-based
encryption scheme called \textsc{Helen}, proposed by Duc and Vaudenay \cite{Helen}. The Lapin protocol is based on an \textprob{LPN} variant called \textprob{Ring-LPN}, where the samples are elements of a polynomial ring.

Khoureich proposed in \cite{hhb} a new version of HB called \textit{h}HB (\textit{harder HB}), which aims to repair susceptibility against GRS attacks. In this paper, we show that this is not the case.

\section{The \textit{h}HB protocol}
The author of \cite{hhb} argues that the weakness of HB is due to that the secret $x$ does not change over time. The \textit{h}HB protocol is built upon the hypothesis; by employing a way of (presumably) securely transmitting a secret session value and letting $x$ take this value, \cite{hhb} aims to patch this weakness. Once the session value has been transmitted, the verifier in the \textit{h}HB protocol runs the standard HB protocol to verify the tag. The \textit{h}HB protocol is outlined in Protocol 1.

\begin{figure}[h*]
			\parbox{\linewidth}{
				\textbf{Protocol 1} {\textbf{(}\textsf{\textit{h}HB protocol}\textbf{)}\\}
			}\hrule height 1pt
			\vspace{0.2cm}~\\
			\parbox[t]{\linewidth}{
			\centering
$\begin{array}{@{}l@{}l@{}c@{}r@{}}
&\textnormal{\bf{Tag}}(s,y) &\qquad\qquad\qquad& \textnormal{\bf{Reader}} (s,y)\vspace{0.2cm}\\ \hline
&\hspace{2.5cm}&\hspace{1.5cm}&\tau \randassign \{0,1\},\xi_0 \randassign \{0,1\}, \xi_1 \randassign \{0,1\}  \\
&(\tau, \xi_0, \xi_1) \leftarrow f_s^{-1}(\alpha, \beta, \gamma, 0^{|s|})&\xleftarrow{\phantom{~}\textstyle (\alpha, \beta, \gamma)\phantom{~}}& (\alpha, \beta, \gamma) \leftarrow f_s(\tau, \xi_0, \xi_1, 0^{|s|}) \\
& \theta \leftarrow \xi_\tau &&  \theta \leftarrow \xi_\tau\\
& p_0 \leftarrow \theta^{|s|} &&  p_0 \leftarrow \theta^{|s|} \\
\hline 
&\textnormal{(Repeat $k$ times)}\\
&&\hspace{1.5cm}&\tau \randassign \{0,1\},\xi_0 \randassign \{0,1\}, \xi_1 \randassign \{0,1\}  \\
&(\tau, \xi_0, \xi_1) \leftarrow f_s^{-1}(\alpha, \beta, \gamma, p_{i-1})&\xleftarrow{\phantom{~}\textstyle (\alpha, \beta, \gamma)\phantom{~}}& (\alpha, \beta, \gamma) \leftarrow f_s(\tau, \xi_0, \xi_1, p_{i-1}) \\
& x_i \leftarrow \xi_\tau &&  x_i \leftarrow \xi_\tau\\
& p_{i-1} \leftarrow  x_1x_2\ldots (x_i)^{|s|-i+1} && p_{i-1} \leftarrow  x_1x_2\ldots (x_i)^{|s|-i+1}\\
\hline 
&\textnormal{(Repeat $r$ times)}\\
& x \leftarrow x_1x_2\ldots x_k && x \leftarrow x_1x_2\ldots x_k\\
& b \randassign \{0,1\}^k & \xrightarrow{\phantom{~~~~~~}\textstyle b\phantom{~~~~~~}}\\
&& \xleftarrow{\phantom{~~~~~~}\textstyle a\phantom{~~~~~~}}&  a \randassign \{0,1\}^k \\
& \nu \leftarrow \BerDist{\epsilon} \\
& z \leftarrow a \cdot x \oplus b \cdot y \oplus \nu  & \xrightarrow{\phantom{~~~~~~}\textstyle z\phantom{~~~~~~}} & \textnormal{Verify } a \cdot x \oplus b \cdot y = z\\

\end{array}$
			}\vspace{0.2cm}\\
			\hrule height 1pt
\end{figure}
The function used by the reader to transmit session values $\tau, \xi_0, \xi_1$,
\begin{equation}
f_s(\lambda_1, \lambda_2, \lambda_3, p_i) \rightarrow (\alpha, \beta, \gamma)
\end{equation}
is defined in Algorithm 1. Similarly, the inverse function used by the tag to decode session values $\tau, \xi_0, \xi_1$,
\begin{equation}
f_s^{-1}(\alpha, \beta, \gamma) \rightarrow (\lambda_1, \lambda_2, \lambda_3, p_i)
\end{equation}
is given in Algorithm 2.
\begin{figure}[h*]
\algsplit{function $f_s$}{alg:f}       
{$\lambda_1, \lambda_2, \lambda_3 \in \{ 0,1\},$\\$ p_i \in \{0,1\}^k$}
{Triple $(\alpha, \beta, \gamma)$}
{ 
    $c_1 \randassign \{0,1\}^k, t_1 \leftarrow c_1 \cdot (s \oplus p_i) \oplus \lambda_1$\\
        $c_2 \randassign \{0,1\}^k, t_2 \leftarrow c_2 \cdot (s \oplus p_i) \oplus \lambda_2$\\
        $c_3 \randassign \{0,1\}^k, t_3 \leftarrow c_3 \cdot (s \oplus p_i) \oplus \lambda_3$\\
        \If{$\lambda_1 \oplus \lambda_2 \oplus \lambda_3 = 0$}{\KwRet{$((c_3,t_3),(c_1,t_1),(c_2,t_2))$}}
        \Else{\KwRet{$((c_2,t_2),(c_3,t_3),(c_1,t_1))$}}
}
\end{figure}
\begin{figure}[h*]
\algsplit{function $f_s^{-1}$}{alg:finv}       
{Triple $(\alpha, \beta, \gamma)$}
{$\lambda_1, \lambda_2, \lambda_3 \in \{ 0,1\},$}
{
        $(c_1, t_1) \leftarrow \alpha, (c_2, t_2) \leftarrow \beta, (c_3, t_3) \leftarrow \gamma$\\
        $\lambda_1 \leftarrow c_1 \cdot (s \oplus p_i) \oplus t_1$\\
        $\lambda_2 \leftarrow c_2 \cdot (s \oplus p_i) \oplus t_2$\\
        $\lambda_3 \leftarrow c_3 \cdot (s \oplus p_i) \oplus t_3$\\
        \If{$\lambda_1 \oplus \lambda_2 \oplus \lambda_3 = 0$}{\KwRet{$(\lambda_2,\lambda_3,\lambda_1)$}}
        \Else{\KwRet{$(\lambda_3,\lambda_1,\lambda_2)$}}
}
\end{figure}

\section{The attack}
We will now proceed with describing our attack. We use a method very similiar to that of \cite{Gilbert05}.
\subsection{Determining secret $y$}
The tag sends the following
$$
    a\cdot x \oplus b \cdot y \oplus  \nu  = z
$$
The verifier checks if the following is satisfied
$$
    a\cdot x \oplus b \cdot y  =^? z,
$$
which is expected to be true for $r \cdot \e{\nu = 0} = r \cdot \epsilon $ of the $r$ samples. If the number of correct equations are above some threshold, the verifier \textsf{accept}s. Otherwise the verifier \textsf{reject}s.

First, we ignore the $x$ vector. By intercepting the communication between the tag and the verifier, we are able to perturbe the interchanged bits. To determine the value of bit of $y$ at index $i$, we run the following steps.

\begin{enumerate}
    \item Let the tag and verifier exchange the session value. For now, this is ignored.
    \item When the tag sends $b$, we flip the $i$th bit. So, $$b_i' \leftarrow b_i \oplus 1.$$
    \item Then, we let the tag and verifier run the $r$ steps. If the reader returns \textsf{accept}, then the bit $y_i$ is very likely to be 0. Naturally, we may amplify the probability of a correct guess by re-running the procedure for the same bit $b_i$.
    \item By repeating for all $k$ bits, we can determine the secret value $y$.
\end{enumerate}
\subsection{Determining secret $s$}
The second stage of the attack aims to determine the secret vector $s$. In the very first step of exchanging the session value $p_0$ is always $0^{|s]}$, which is the key to our exploit. To determine the value of bit of $s$ at index $j$, we run the following steps.
\begin{enumerate}
\item In the first step of the session-value exchange, flip the $j$th bit in $c_1$. As a result, we have
\begin{equation}
    \lambda_1 \leftarrow (c_1 \oplus \delta_j)\cdot s \oplus t_1.
\end{equation}
where $\delta_j$ is a vector with 1 on index $j$ and all-zero on the remaing indices.
Hence,
\[ \lambda_1 \leftarrow  \begin{cases} \lambda_1 & \quad \text{if } s_j = 0,\\ \lambda_1 \oplus 1 & \quad \text{if } s_j = 1.\\ \end{cases} \]
Applying the same procedure to $c_2$ and $c_3$, we are able to conditionally flip also $\lambda_2$ and $\lambda_3$.
\item If the two values satisfy $\xi_0 = \xi_1$ (which is true with probability $\frac12$), then $x$ will be perturbed at position 0, i.e., $x_0 \leftarrow x_0 \oplus 1$. Everything else remains the same.
\item When the tag is verfied against the reader, we set the $j$th bit of $a$ to always be 1. Hence, the tag will compute 
\begin{equation}
    z \leftarrow \begin{cases} a \cdot x \oplus b \cdot y \oplus \nu & \quad \text{if } s_j = 0,\\ a \cdot x \oplus b \cdot y \oplus \nu \oplus \phi & \quad \text{if } s_j = 1.\\ \end{cases} 
\end{equation}
where $\pr{\phi = 1} = \frac12$. 
So, if $s_j = 1$, then the reader will output $\textsf{reject}$ with probability $\frac12$. Running the procedure a polynomial number of times for the same index $j$ will give a good estimate of $s_j$.
\end{enumerate}
Implementation of hHB and the MITM attack can be found at \cite{hhb_source}.
%
%
%
%
%
%
%
\bibliographystyle{plain}
\bibliography{cryptogroup,dissertation}
\end{document}